# Background Proportional Enhancement of the Extended Fine Structure in Nonresonant Inelastic X-ray Scattering

T.T. Fister[1], G.T. Seidler[1,(*)], C. Hamner[1], J.O. Cross[2], J.A. Soininen[3], J.J. Rehr[1]

1. Physics Department, University of Washington, Seattle, Washington 98105
2. Argonne National Laboratory, Argonne, IL 60439
3. Division of X-ray Physics, Department of Physical Sciences, University of Helsinki, FIN-00014 Finland
*corresponding author: seidler@phys.washington.edu

**Abstract.** We report new measurements and calculations of the non-resonant inelastic x-ray scattering (NRIXS) from Mg and Al for a wide range of momentum transfers, $q$. Extended oscillations in the dynamic structure factor $S(q,\omega)$ due to scattering from the 2$p$ and 2$s$ orbitals (i.e. $L$-edges) are observed out to more than 150 eV past the binding energy. These results are discussed in context of the recently proposed representation of $S(q,\omega)$ for core shells as an atomic background modulated by interference between different photoelectron scattering paths, in analogy to the standard treatment of extended x-ray absorption fine structure. In agreement with this representation, we find a strong increase in the atomic background with increasing $q$ with a concomitant enhancement in the amplitude of the extended fine structure. This effect should be generic and hence may enable improved measurement of the extended fine structure in a wide range of materials containing low-Z elements.

## INTRODUCTION

The non-resonant inelastic x-ray scattering from core electrons (NRIXS) has often been treated as a bulk alternative to extended x-ray absorption fine structure (XAFS) for low-Z materials[1-3] For small momentum transfer $q$, this is a natural comparison since both techniques measure the same excited final states, specifically those which can be reached by a dipole transition. On the other hand, NRIXS measurements at higher $q$ can access new final states through non-dipolar transitions, uniquely providing a more complete perspective of electronic excitations in the system.[4,5]

This promise of additional information contained in $q$-dependent NRIXS has been tempered by two factors. First, the small cross-section for core-shell NRIXS has made it difficult to observe extended fine structure, except in a few cases[1,6]. Second, there has been a lack of theoretical interpretation for core-shell NRIXS when far past the edge, where current methods using the Bethe-Salpeter equation[7] become numerically restrictive. Considerable progress has recently been made on both issues. The development of multicrystal NRIXS spectrometers has shortened experimental measurement times[8], and the real-space multiple scattering (RSMS) approach frequently used in calculations of XAFS spectra has been extended to treat $q$-dependent NRIXS.[9]

We report here new measurements of the NRIXS extended fine structure for the 2$p$ and 2$s$ orbitals of Mg and Al. These materials are of interest in the present context due to the need for careful consideration of the interaction between the core-hole and the photoelectron, and also because their atomic physics requires the presence of a rich final state spectrum including substantial contributions with different angular momentum. Hence, these materials are well-suited for detailed testing of the RSMS approach to $q$-dependent NRIXS.

We find generally good agreement between theory and experiment, and in particular verify one prediction which may have considerable impact for many future studies: we find a signal enhancement of the extended fine structure which is proportional to the strong $q$-dependent evolution in shape and magnitude of the atomic background. This enhancement allows us to easily resolve extended fine structure which is quite small in direct soft x-ray XAFS measurements of the same low-energy edges. This effect should be generic, suggesting that NRIXS measurements outside the dipole limit may be valued not only for their sensitivity to a rich spectrum of final states in the near-edge region, but also for the signal enhancement provided for extended oscillations in this regime.

Below, we first review the theory governing NRIXS, especially including the $q$-dependent RSMS approach. We next discuss experimental details. The experimental and theoretical results are then presented and discussed, with emphasis on the limitations of the quasiparticle treatment of the photoelectron in the RSMS implementation and on the challenges for both theory and experiment which are provided by the large background from the valence Compton scattering. We then conclude.

## THEORY

The double differential cross section for NRIXS can be expressed in terms of the dynamic structure factor, $S(q,\omega)$, as[10]

$$\frac{d^2\sigma}{d\omega d\Omega} = \left(\frac{d\sigma}{d\Omega}\right)_{Th} S(\mathbf{q},\omega)$$

$$= \left(\frac{d\sigma}{d\Omega}\right)_{Th} \sum_f \left|\langle f | \sum_j e^{i\mathbf{q}\cdot\mathbf{r}_j} | i \rangle\right|^2 \delta(E_f - E_i - \hbar\omega). \quad (1)$$

In Equation (1), $(d\sigma/d\Omega)_{Th}$ is the Thomson differential cross-section, $j$ indexes the electrons' coordinates, and the subscripts $i$ and $f$ refer to the initial and final states respectively.

When $qa$ is small, where $a$ is the average radius of the initial state, only the first non-zero term in $e^{i\mathbf{q}\cdot\mathbf{r}}$ significantly contributes to $S(q,\omega)$. This restricts final states to transitions that satisfy the angular momentum $l$ selection rule $\Delta l = \pm 1$, i.e., only dipole transitions are present for $qa \ll 1$. In this limit, it is well known that XAFS and XRS are sensitive to the same transition matrix element, except with the direction of $q$ corresponding to the polarization vector of incident radiation in XAFS[3,5]. Electron energy loss spectroscopy (EELS) also measures $S(q,\omega)$, but is largely limited to dipole transitions due to the $q^{-4}$ dependence in $(d\sigma/d\Omega)_{Th}$ for electron scattering.

At higher $qa$, NRIXS can provide different information from XAFS because of transitions to dipole-forbidden final states. For polycrystalline materials, it has recently been shown[9] that

$$S(q,\omega) = \sum_l S_l(q,\omega)$$
$$= \sum_l (2l+1)|M_l(q,\omega)|^2 \rho_l(\omega), \quad (2)$$

where $S_l(q,\omega)$ is the contribution to the dynamic structure factor from final states with angular momentum $l$, $M_l(q,\omega)$ is a transition weighting factor that only depends on the initial state wave function[9,11], and $\rho_l$ is the unoccupied density of states projected onto an angular momentum basis, i.e. the $l$DOS.

The question then arises as to how to calculate $\rho_l$ for a given system. Soininen et al.[9] have expanded FEFF, a software package which has seen extensive use to calculating XAFS and related dipole-limited optical transitions[12], to include higher order, multipole transitions and thus to calculate $S(q,\omega)$ for NRIXS using Equation (2). The key step in calculating $\rho_l$ is the numerical determination of the photoelectron's real-space propagator, which is expressed as the sum of contributions from the potential of the source atom and from all photoelectron scattering paths.

The clean distinction between atomic and scattering contributions in the photoelectron propagator is well-known in calculations of the x-ray absorption coefficient $\mu$[12] and leads to the standard expression

$$\mu(\omega) = \mu_0(\omega)[1 + \chi(\omega)], \quad (3)$$

where $\mu_0$ is referred to as the atomic background and $\chi(\omega)$ contains all information about interference among the different photoelectron scattering paths. Likewise, it has recently been shown that the dynamic structure can be expressed as[9]

$$S(q,\omega) = S_0(q,\omega)[1 + \chi(q,\omega)]. \quad (4)$$

The separation of $S(q,\omega)$ into atomic and interference contributions (i.e., $S_0(q,\omega)$ and $\chi(q,\omega)$, respectively) is conceptually important because $S_0(q,\omega)$ depends strongly on $q$, with its peak intensity moving to higher energies as $q$ increases. This has a nontrivial experimental consequence: extended oscillations in $\chi$ should be more easily observed at high $q$ because of the change in shape of $S_0(q,\omega)$. Furthermore, at appropriately high $q$, extended fine structure should be easily measurable in $S(q,\omega)$ at high $q$ that would have very small amplitude in the analogous soft x-ray XAFS experiment because of the rapid $\omega^{-3}$ decrease of $\mu_0(\omega)$ in the extended regime. The extended fine structure is yet more strongly suppressed in EELS, which has an $\omega^{-4}$ decrease in its atomic background.[13]

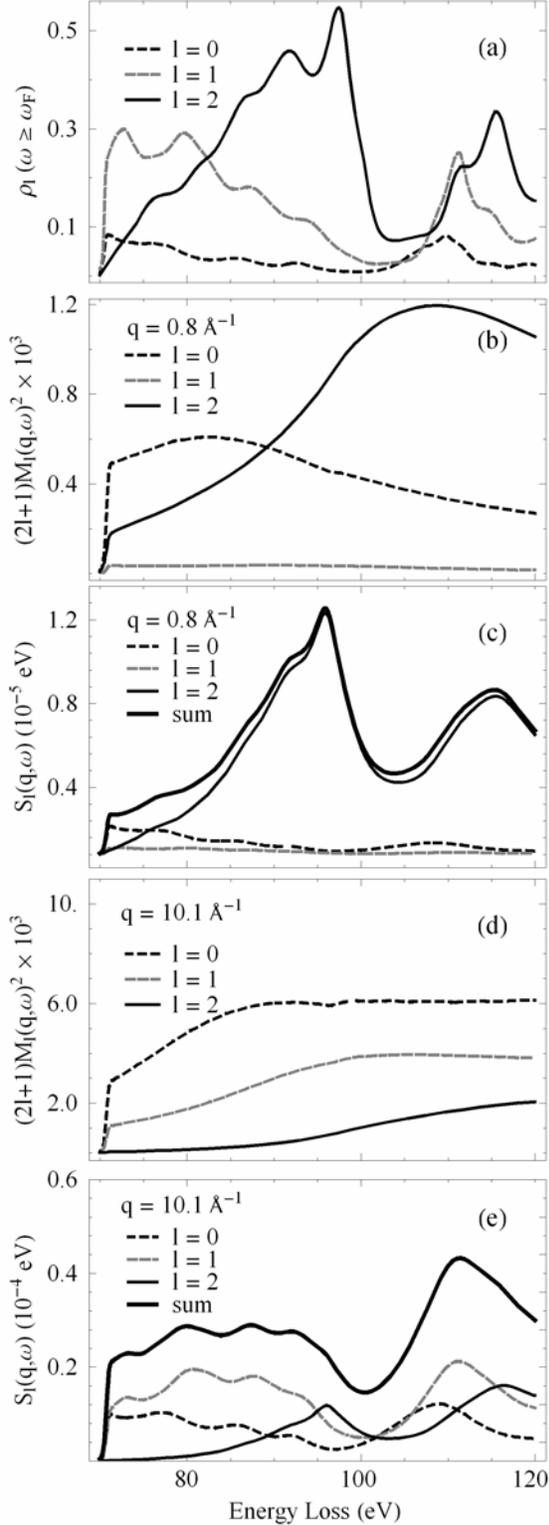

**FIG. 1**: Calculations for the $L_2$-edge of Al. (a) The final density of states projected onto the angular momentum basis, i.e., the $l$DOS, (b) the $l$DOS weighting factor (see Equation (2)) when $q = 0.8$ Å$^{-1}$, (c) $S(q,\omega)$ and $S_l(q,\omega)$ when $q = 0.8$ Å$^{-1}$, (b) the $l$DOS weighting factor when $q = 10.1$ Å$^{-1}$, (c) $S(q,\omega)$ and $S_l(q,\omega)$ when $q = 10.1$ Å$^{-1}$.

By means of orientation, it is useful to consider representative calculations the quantities appearing in Equations (2) and (4). We focus on the $L_2$-edge of polycrystalline Al as it will be relevant for the subsequent interpretation of experimental results. Recall that the $L_2$-edge occurs at 73 eV and corresponds to the $2p_{1/2}$ initial state with a 0.34 Å$^{-1}$ average radius.

From top to bottom in Fig. 1, we show the calculated $l$DOS for the Al $L_2$ edge, the weighting factor $(2l+1)|M_l(q,\omega)|^2$ and $S_l(q,\omega)$ for $q = 0.8$ Å$^{-1}$, and the weighting factor and $S_l(q,\omega)$ for $q = 10.1$ Å$^{-1}$. These $q$ are the two extremes in the experiment and correspond to $qa = 0.27$ and 3.4, respectively. We used the full multiple scattering algorithm (FMS) for a cluster of 134 atoms and did not include Debye-Waller effects in the calculation. The low-$q$ $S(q,\omega)$ (Fig. 1, part (c)) is dominated by $p \to d$ transitions, whereas the high-$q$ $S(q,\omega)$ (Fig 1, part (e)) has additional multipole channels leading to strong contributions from each of the $s, p,$ and $d$ channels. Each calculation of $S(q,\omega)$ uses the same $l$DOS, but the different limiting cases of $q$ have dramatically different $|M_l(q,\omega)|^2$ coefficients, as shown in parts (b) and (d) of Fig 1.

Moving to a wider energy range, we show in Fig. 2 Al $L_2$-edge calculations for both $S_0(q,\omega)$ (gray) and $S(q,\omega)$ (black) for $q$ ranging from 0.8 to 10.1 Å$^{-1}$, again taking the same values as in the experiment. For the same energy range and momentum transfers, we show $\chi(q,\omega)$ in Fig. 3. Vertical, dashed lines are visual guides and are at the same energies in each figure. FMS calculations in the near-edge region have been merged with path-expansion results used in the extended regime. The lack of strong $q$-dependence for the extended fine-structure in $\chi(q,\omega)$ is due to the strong admixture of final state symmetries. Atleast in systems lacking strong anisotropy, we expect this will be a generic effect.

One key detail in Figs.2 and 3 deserves special attention. Note the steady shift of the peak of $S_0(q,\omega)$ to higher energy with increasing $q$ and the concomitant *proportional* increase in the predicted amplitude of the fine structure. As a result, one expects that the peaks at 160 and 210 eV will be easier to resolve outside the dipole limit, even though the amplitude of $\chi(q,\omega)$ is insensitive to $q$ (Fig. 3).

It is a central goal of this study to test this prediction. The enhancement of the observability of the extended fine structure should be a generic effect, and may therefore have broad application in improving

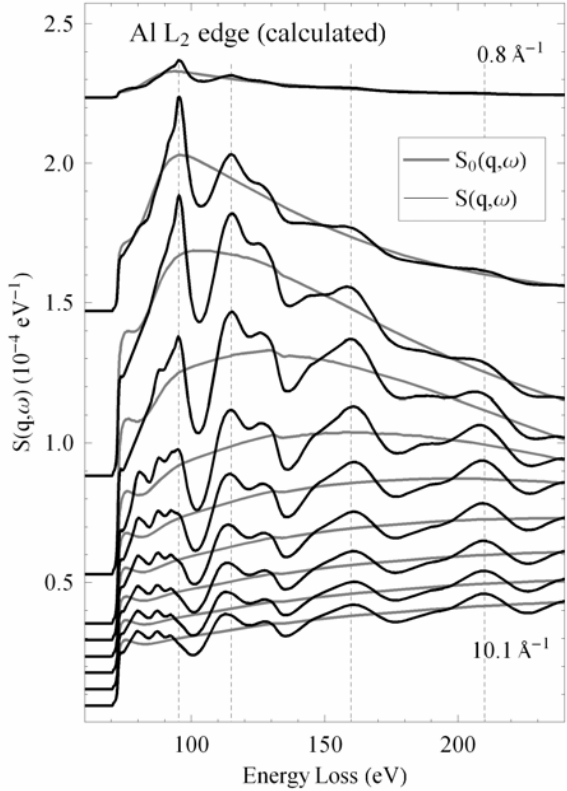

**FIG. 2**: Calculations of the atomic ($S_0(q,\omega)$, gray curves) and total ($S(q,\omega)$, black curves) contribution to the dynamic structure factor for the Al $L_2$-edge at the ten momentum transfers used in the experiment. Each curve has been offset for clarity. The dashed lines are guides to the eye.

NRIXS and the corresponding x-ray absorption spectroscopy studies.

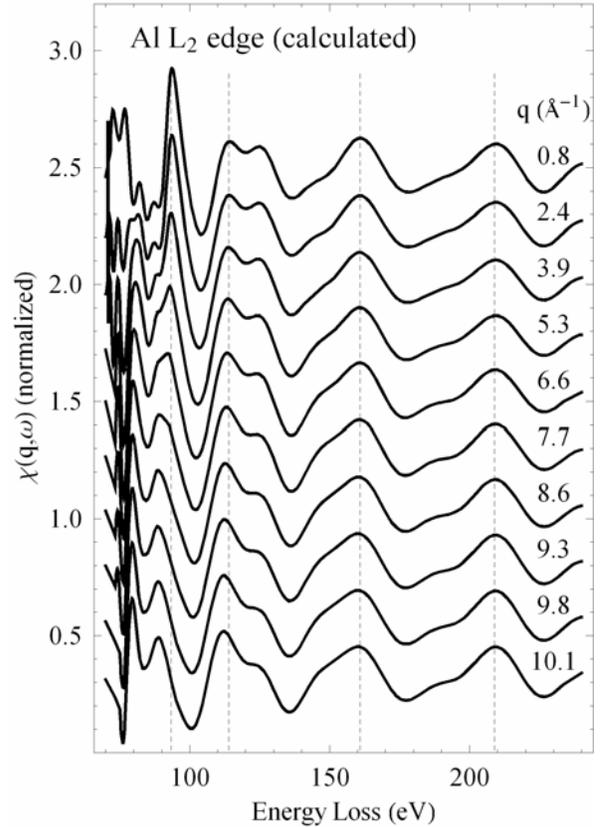

**FIG. 3**: Calculation for the fine structure $\chi(q,\omega)$ for the Al $L_2$-edge for the ten momentum transfers used in the experiment. Each curve has been offset for clarity but is otherwise on the same absolute scale. The dashed lines are guides to the eye, and are at the same energies as in Fig. 2.

## EXPERIMENT

We performed NRIXS measurements at 20-ID, an undulator beamline operated by the Pacific Northwest Consortium X-ray Operations Research sector at the Advanced Photon Source. We made use of the new lower energy inelastic x-ray scattering (LERIX) spectrometer,[8] which permits simultaneous measurement of NRIXS energy loss-spectra at multiple values of $q$. Working at the Si (555) reflection from the LERIX analyzers with ~10 keV incident photons, $q \approx$ 0.8, 2.4, 3.9, 5.3, 6.6, 7.7, 8.6, 9.3, 9.8, 10.1 Å$^{-1}$. The incident flux was ~$5 \times 10^{12}$ photons/s and the overall energy resolution was monochromator-limited to 1.3 eV. We used 1 cm$^2$ Al and Mg foil samples (99.9% purity) that were one absorption length thick at 10 keV. The samples were oriented so that the beam was normal to the face of the foil, ~200 $\mu$m below the top edge. For both samples, measurements were performed both at room temperature (300 K) and also at 100K using a nitrogen flow cryostat. Temperature stability in the flow cryostat was ~ 5 K during the measurements. For the room temperature measurements the samples were in a He environment; for the 100 K measurements the samples were in vacuum.

Calibration of the energy loss scale is better than 0.1 eV[8]. The energy resolution is 1.3 eV, which is the theoretical limit for the monochromator. We corrected for small (< 0.05 eV), scan-to-scan shifts in the monochromator energy by aligning each scan's edge energy. The data from successive scans is binned and assigned a statistical uncertainty from Poisson statistics. We also corrected for the approximately linear energy dependence due to sample absorption and the $\omega_2/\omega_1$ contribution to the Thomson prefactor in Equation (1). The data was normalized according to the f-sum rule[14,15]

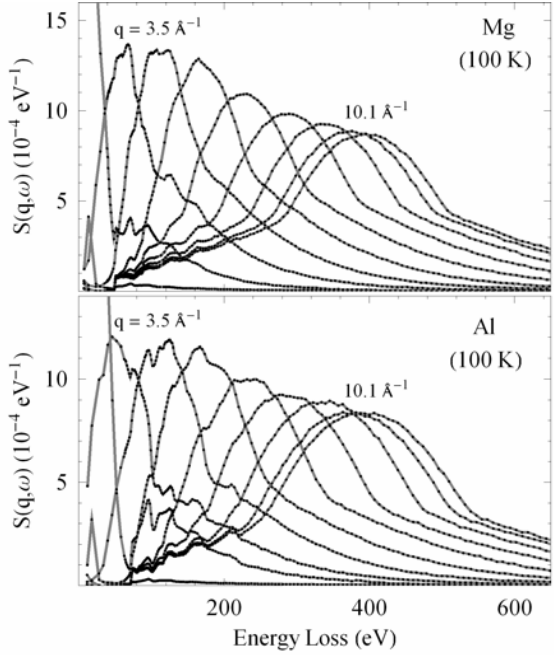

**FIG. 4:** Mg (top panel) and Al (bottom panel) NRIXS measured at 100 K.

$$\int S(q,\omega)\omega\,d\omega = N\frac{\hbar^2 q^2}{2m}, \quad (5)$$

where $N$ is the number of electrons contributing to $S(q,\omega)$. This expression intuitively confirms that the average value for $\omega$ is the Compton shift $\hbar^2 q^2/2$, as expected in the impulse approximation[10]. We normalized the 0.8 and 1.2 Å$^{-1}$ spectra using just the valence contribution and the higher $q$ data with all but the $K$-electron contribution. In the latter case, we accounted for the ~10% change in $q$ from the increasing energy loss by normalizing the spectrum to the momentum transfer at the center of the Compton profile. The overall magnitude of the core-shell contribution to the normalized spectra were generally within 10% of the corresponding theoretical calculations.

## RESULTS AND DISCUSSION

In Fig. 4, we present NRIXS measurements for Mg (top panel) and Al (bottom panel) at 100 K. In each case, the statistical uncertainty is comparable to the size of each symbol or smaller. With increasing $q$, the systematic broadening and energy shift of the valence Compton scattering is apparent. Also clear is the presence of considerable near-edge and extended fine structure due the NRXIS from the 2$p$ and the 2$s$ orbitals in each material. Note that the binding energy for the 2$p$ orbital ($L_{2,3}$-edge) is 50 eV for Mg and 73 eV for Al, while the binding energy for 2$s$ orbital ($L_1$-edge) is 88 eV for Mg and 118 eV for Al.

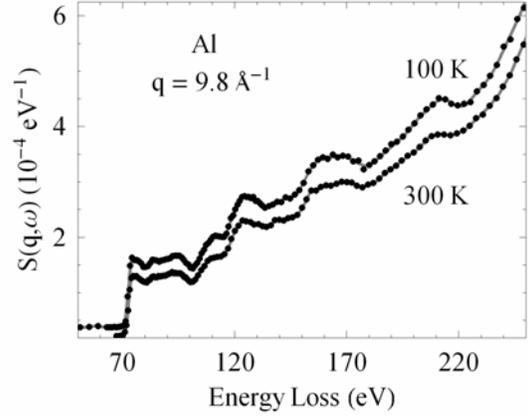

**FIG. 5:** A comparison of $S(q,\omega)$ at $q$ = 9.8 Å$^{-1}$ for Al at T = 100 K and 300 K. Note the increased amplitude of the extended oscillations upon cooling.

In Fig. 5, we compare NRIXS at room temperature and 100 K for $q$ = 9.8 Å$^{-1}$ for Al. The only significant difference in the spectra is the decrease in amplitude of the extended oscillations at room temperature, as expected from the thermal Debye-Waller factor of

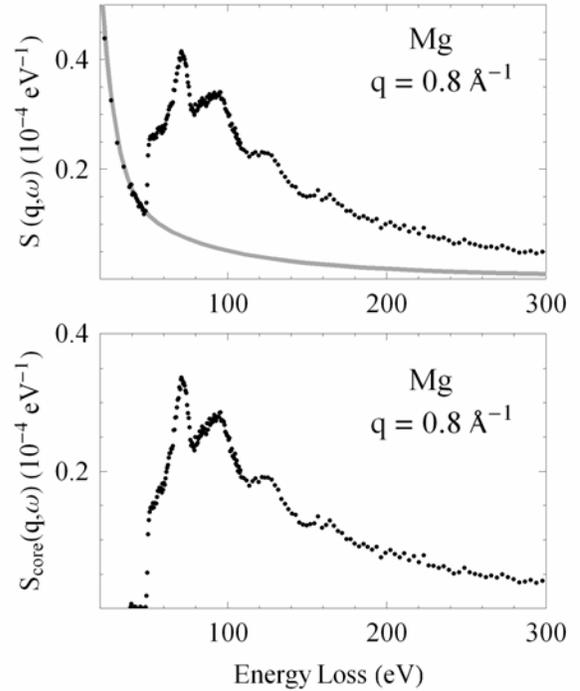

**FIG. 6:** $S(q,\omega)$ for Mg when $q$ = 0.8 Å$^{-1}$, before (top) and after (bottom) subtracting an *ad hoc* fit to the high-energy tail of the valence Compton scattering.

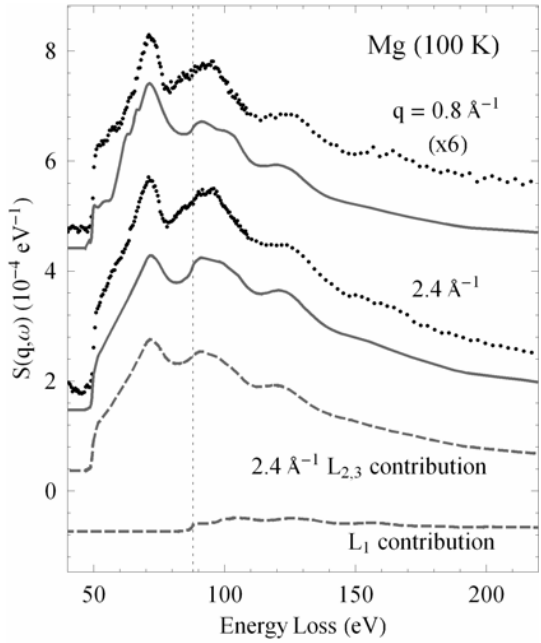
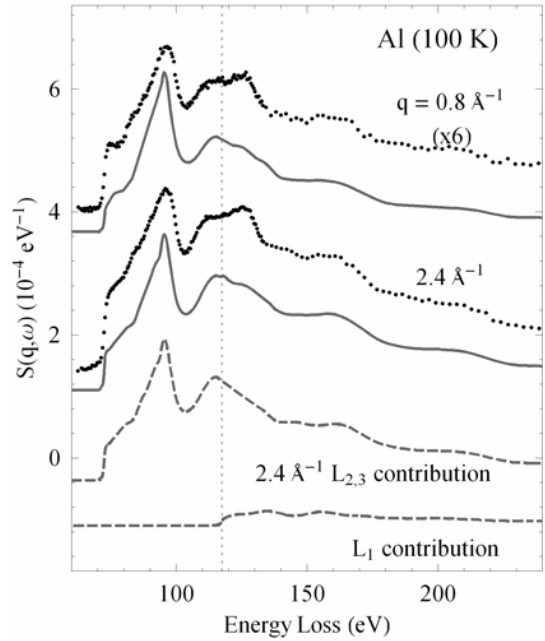

**FIG. 7:** Measurement (points) and calculation (solid lines) of the contribution to $S(q,\omega)$ from the Mg $L$-edges at $q = 0.8$ Å$^{-1}$ and 2.4 Å$^{-1}$. The 0.8 Å$^{-1}$ curves have been scaled by a factor of 6 to aid comparison. The calculated contributions from each edge for $q = 2.4$ Å$^{-1}$ are also shown (dashed curves). All curves have been offset for clarity. The vertical dashed line indicates the position of the $L_1$-edge.

**FIG. 8:** Measurement (points) and calculation (solid lines) of the contribution to $S(q,\omega)$ from the Al $L$-edges at $q = 0.8$ Å$^{-1}$ and 2.4 Å$^{-1}$. All presentation details are analogous to Fig. 7.

~400 K for Al[16]. Within the Einstein model[17], the 100 K Debye-Waller factors for both Mg and Al are within 10% of the zero-temperature limit. Hence, we focus only on the lower temperature data to better resolve the extended fine structure. For reference, the room temperature data for Mg has been previously been presented elsewhere[8], but was not analyzed or discussed.

A key issue in the interpretation of the core-shell contribution to NRIXS is the subtraction of the valence Compton background[4,15]. This process is most difficult when the valence Compton scattering is centered near the binding energy of the core contribution, especially for lower binding energies where the valence Compton background is sharply peaked in energy. As a consequence, oscillations in $\chi(q,\omega)$ which are most strongly enhanced by a large value in $S_0(q,\omega)$ may be the most difficult to analyze.

In the long term, the core and valence NRIXS must be treated on the same theoretical basis for the fullest application of this spectroscopy. Unfortunately, such a treatment is not presently available. On the one hand, FEFF8.2[12], which we find here works well for the core NRIXS, has difficulty with valence calculations due to its use of a single atomic initial state and of spherical muffin-tin potentials. On the other hand, successful approaches for calculating the core and valence profile in the impulse limit of extremely high $q$ make use of plane-wave final states and hence are not applicable to determining the NRIXS fine structure.[14,18] Combining the core and valence contributions into *ab initio* $q$-dependent NRIXS calculations will require better simultaneous treatment of the initial and final states[19].

We separate our data into three categories: low-$q$ (0.8 and 2.4 Å$^{-1}$) where the core-shell NRIXS is on the high-energy tail of the valence Compton scattering; intermediate-$q$ (3.9, 5.3, 6.6, and 7.7 Å$^{-1}$) where large portions of the fine structure intractably coincide with the peak of the valence Compton scattering; and high-$q$ (8.6, 9.3, 9.8 and 10.1 Å$^{-1}$) where the valence Compton has largely moved beyond the first 200 eV of fine structure. Given the difficulties described above, we focus exclusively on the low- and high-$q$ regimes.

In Fig. 6 we show an expanded view of $q = 0.8$ Å$^{-1}$ NRIXS for Mg at 100 K. The tail of the valence Compton scattering can be reasonably fit with a Lorentzian function for background subtraction (shown as the gray line in Fig. 6). The remaining low-$q$ $L$-edge spectra which result from this *ad hoc* background subtraction are shown in Figs. 7 and 8 for Mg and Al respectively. The 0.8 Å$^{-1}$ spectrum has been scaled by a factor of six for ease of comparison. It is important to note that our low-$q$ results are in good agreement with previous measurements from soft x-ray XAFS[20] and from EELS[21].

Also shown in Fig. 7 and 8 is comparison with FEFF calculations at each $q$ and a separation of the $L_{2,3}$- and $L_1$-contributions. As expected from Fig. 1, the large quantity of $d$-type final states dominates the spectrum. This can be seen in the data from the ratio of $L_{2,3}$ to $L_1$-edge heights, which is much higher than the 3:1 ratio of initial state electrons due to low-$q$ dipole transitions from from $L_{2,3}$ $2p$ initial states to $d$-type final states that far outweigh the transitions from $L_1$ $2s$ initial state to $p$-type final states.

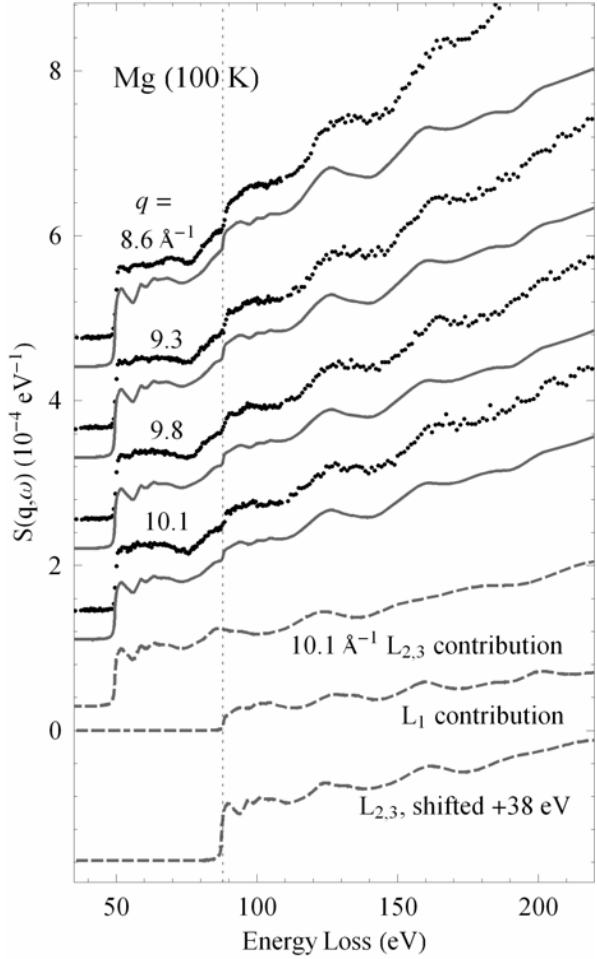

**FIG. 9:** Measurement (points) and calculation (solid curves) of $S(q,\omega)$ for Mg $q$ = 8.6, 9.3, 9.8, and 10.1 Å$^{-1}$. The calculated contributions from each edge for $q$ = 10.1 Å$^{-1}$ are also shown (dashed curves). All curves have been offset for clarity. The vertical dashed line indicates the position of the $L_1$-edge.

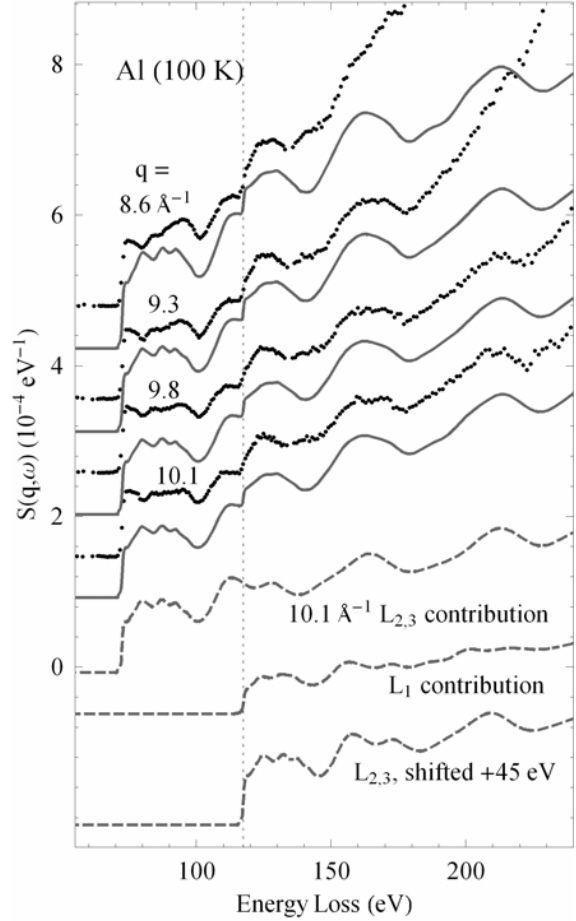

**FIG. 10:** Measurement (points) and calculation (solid curves) of $S(q,\omega)$ for Al $q$ = 8.6, 9.3, 9.8, and 10.1 Å$^{-1}$. The calculated contributions from each edge for $q$ = 10.1 Å$^{-1}$ are also shown (dashed curves). All curves have been offset for clarity. The vertical dashed line indicates the position of the $L_1$-edge.

In the calculations, full multiple scattering[22] (FMS) was used in the first 40-50 eV for each edge, giving way to a path expansion calculation[23] for higher photoelectron energies. Convergence in both cases requires a 100 atom cluster and a maximum $l$ (see Equation (2)) of two for the near-edge, FMS calculation and a maximum $l$ of 10 for the extended fine structure's path expansion calculation. The energy axis was shifted by 2 eV in the Mg $L_{2,3}$ FMS calculation to match the prominent feature at 71 eV. A shift of this nature is not uncommon in XAFS analysis and is likely due to shortcomings in the calculation of exchange and correlation effects.[24] Spectral features past the edge are shifted when the photoelectron is treated as a quasiparticle. In the extended regime, these shifts approach a constant value. In the near-edge region, however, these quasiparticle corrections vary more strongly with energy. This type of discrepancy is apparent in the Mg results shown in Fig. 7: aligning the spectrum to the 71 eV peak results in poor agreement with the position of the second significant peak in the $L_{2,3}$ spectrum that occurs at ~91 eV in the theoretical calculation and at ~83 eV in the data. This feature is obscured by the $L_1$-edge, but is more prominent at higher $q$ where we have used a different correction for the calculated Mg $L_{2,3}$ edge. Shown in Fig. 8, the analogous $L_{2,3}$ calculation for Al did not require a shift to align the first major feature at 96 eV; nonetheless, the calculation did exhibit the misalignment of the first two significant features occurring at 96 and 114 eV, as was the case in the Mg data.

Moving on to the high-$q$ regime, we present high-$q$ NRIXS (without background subtraction) for the Mg and Al $L$-edges in Figs. 9 and 10 respectively. Note the dramatic enhancement in the amplitude of the extended oscillations for the high $q$ measurements in Figs. 9 and 10 as compared to the corresponding low $q$ measurements in Figs. 7 and 8, respectively. Hence, we do observe the predicted signal enhancement in the extended fine structure.

Also note the increase in the relative strength of the $L_1$-edge relative to the low-$q$ case. This is due to the proportional increase in $S_0(q,\omega)$ near the $L_1$-edge energy at high $q$. Additionally, we show the calculated $L_{2,3}$-edge shifted to the $L_1$-edge position to highlight the similarity in their fine structure when dipole *and* quadrupole transitions access common *s*-, *p*-, and *d*-final states from each edge. As shown in part (e) of Fig. 1, the contribution to the dynamic structure factor from each $S_l(q,\omega)$ is comparable in the high-$q$ limit, leading to a final state spectrum that is roughly independent of initial state symmetry.

We used the same theoretical parameters for the low- and high-$q$ results, with the exception of Mg's $L_{2,3}$ energy shift, which was reversed to -2 eV for better agreement with the feature coinciding with the $L_1$-edge. With the extended fine structure and the $L_1$-edge better-resolved, we shifted the extended $L_{2,3}$ calculations by +6 eV and +3 eV for the Mg and Al calculations respectively. The calculation for the Mg $L_1$-edge was shifted by -2 eV in the FMS regime and by +1 eV for the extended fine structure. The Al $L_1$-edge had analogous shifts of -1 and +1 eV. Finally, a small exponential background was added to the calculations in the high-$q$ calculations as an *ad hoc* contribution from the low energy tail of the valence Compton profile. The agreement between theory and experiment is again generally good, allowing for the shortcomings of the treatment of the quasiparticle corrections, as we have been careful to identify in this discussion.

## CONCLUSIONS

In summary, we present new measurements and calculations of the nonresonant inelastic scattering from Mg and Al over a wide range of momentum transfers. These results demonstrate a strong transition from a dipole to non-dipole scattering limit for the contribution from the 2*s* and 2*p* initial states to the scattering. On increasing momentum transfer, we observe that the evolution of the atomic contribution to the dynamic structure function results in a large enhancement of amplitude of the extended fine structure relative to measurements of x-ray absorption spectra. This effect should be generic, and may provide a route for generally improved measurements of extended fine structure for materials containing high concentrations of low-Z elements. The experimental results and real-space multiple scattering theory are in generally good agreement, but demonstrate the importance of further improvements in theoretical treatments of the photoelectron self-energy.

## ACKNOWLEDGMENTS

This research was supported by DOE, Basic Energy Science, Office of Science, Contract Nos. DE-FGE03-97ER45628 and W-31-109-ENG-38, ONR Grant No. N00014-05-1-0843, the Summer Research Institute program at the Pacific Northwest National Lab, and the Academy of Finland, Contracts No. 201291, 205967, and 110571. The operation of Sector 20 PNC-CAT/XOR is supported by DOE Basic Energy Science, Office of Science, Contract No. DE-FG03-97ER45629, the University of Washington, and grants from the Natural Sciences and Engineering Research Council of Canada. Use of the Advanced Photon Source was supported by the U.S. Department of Energy, Basic Energy Sciences, Office of Science, under Contract W-31-109-Eng-38. We thank Ed Stern, Yejun Feng, Kevin Jorrisen and Tim Elam for helpful discussions in all phases of the experiment. We wish to thank Micah Prange and Josh Kas for assistance with the *q*-dependent FEFF code and interpretation of the results.